\PassOptionsToPackage{dvipsnames}{xcolor}
\documentclass[sigconf]{acmart}
\usepackage[utf8]{inputenc}
\usepackage{hyperref}
\usepackage{graphicx}
\usepackage{enumerate}
\usepackage[framemethod=TikZ]{mdframed}
\usepackage{array}
\usepackage{multicol}
\usepackage{booktabs}
\usepackage{multirow}
\usepackage{float}
\usepackage{subcaption}
\usepackage{harveyballs}

\newcommand{\secref}[1]{Sec.~\ref{#1}}
\newcommand{\figref}[1]{Fig.~\ref{#1}}
\newcommand{\tabref}[1]{Tab.~\ref{#1}}

\newenvironment{sidebarbox}
  {\mdfsetup{
    innertopmargin=4pt,
    linewidth=0pt,
    frametitleaboveskip=-\ht\strutbox,
    frametitlealignment=\center,
    backgroundcolor=sidebarbgcolor
    }
  \begin{mdframed}%
  \begin{minipage}[t]{\linewidth}
  }
  {\end{minipage}%
  \end{mdframed}}

\newcommand{\smallSE}[1][0.9ex]{%
  \harveyInternal[#1]{OliveGreen}{\harveyBallQuarter}
}
\newcommand{\mediumSE}{%
  \harveyInternal{blue}{\harveyBallHalf}
}
\newcommand{\lotSE}{%
  \harveyInternal{Orange}{\harveyBallThreeQuarter}
}
\newcommand{\fullSE}{%
  \harveyInternal{Red}{\harveyBallFull}
}

\newcommand{\harveyInternal}[3][0.9ex]{%
  \def\harveyBallsSize{#1}%
  \def\harveyBallsColor{#2}%
  \def\harveyBallsLineColor{#2}%
  #3
}

\definecolor{sidebarbgcolor}{HTML}{ffb300}
\mdfsetup{%
   backgroundcolor=gray!3,
   roundcorner=3pt%
}

\def\harveyBallsSize{0.5ex}
\def\inlineSize{0.6ex}

\AtBeginDocument{%
  \providecommand\BibTeX{{%
    \normalfont B\kern-0.5em{\scshape i\kern-0.25em b}\kern-0.8em\TeX}}}

\copyrightyear{2024}
\acmYear{2024}
\setcopyright{acmlicensed}\acmConference[MODELS Companion '24]{ACM/IEEE 27th International Conference on Model Driven Engineering Languages and Systems}{September 22--27, 2024}{Linz, Austria}
\acmBooktitle{ACM/IEEE 27th International Conference on Model Driven Engineering Languages and Systems (MODELS Companion '24), September 22--27, 2024, Linz, Austria}
\acmDOI{10.1145/3652620.3688343}
\acmISBN{979-8-4007-0622-6/24/09}

\title{Digital Twin Evolution for Sustainable Smart Ecosystems}

\author{Judith Michael}
\orcid{0000-0002-4999-2544}
\affiliation{
  \institution{RWTH Aachen University}
  \city{Aachen}
  \country{Germany}
}
\email{michael@se-rwth.de}

\author{Istvan David}
\orcid{0000-0002-4870-8433}
\affiliation{
  \institution{McMaster University}
  \city{Hamilton}
  \country{Canada}
}
\email{istvan.david@mcmaster.ca}

\author{Dominik Bork}
\orcid{0000-0001-8259-2297}
\affiliation{
  \institution{TU Wien}
  \city{Vienna}
  \country{Austria}
}
\email{dominik.bork@tuwien.ac.at}

\begin{document}

\emergencystretch 3em

\begin{abstract}
Smart ecosystems are the drivers of modern society. They control infrastructures of socio-techno-economic importance, ensuring their stable and sustainable operation.
Smart ecosystems are governed by digital twins---real-time virtual representations of physical infrastructure. To support the open-ended and reactive traits of smart ecosystems, digital twins need to be able to evolve in reaction to changing conditions.
However, digital twin evolution is challenged by the intertwined nature of physical and software components, and their individual evolution. 
As a consequence, software practitioners find a substantial body of knowledge on software evolution hard to apply in digital twin evolution scenarios and a lack of knowledge on the digital twin evolution itself. 
The aim of this paper, consequently, is to provide software practitioners with tangible leads toward understanding and managing the evolutionary concerns of digital twins. 
We use four distinct digital twin evolution scenarios, contextualized in a citizen energy community case to illustrate the usage of the 7R taxonomy of digital twin evolution. 
By that, we aim to bridge a significant gap in leveraging software engineering practices to develop robust smart ecosystems.
\end{abstract}



\maketitle

\begin{acks}
Funded by the Deutsche Forschungsgemeinschaft (DFG, German Research Foundation) under Germany’s Excellence Strategy -- EXC 2023 Internet of Production -- 390621612. Website: \url{https://www.iop.rwth-aachen.de}.
\end{acks}

\section{Introduction}

Our modern world runs by smart ecosystems---large-scale, decentralized systems, capable of self-organization and self-optimization~\cite{jensen2020applying}.
Examples of smart ecosystems include smart cities~\cite{graciano_neto2023what}, smart energy communities~\cite{gramelsberger2023enabling}, and smart grids with renewable components~\cite{hasan2023review}.

Much like natural ecosystems, smart ecosystems are open-ended and need to allow for continuous changes in their structure and behavior. These evolutionary dynamics, in turn, challenge the technical sustainability~\cite{penzenstadler2018software} of smart ecosystems, i.e., their ability to maintain the quality of service over a prolonged period of time~\cite{hilty2006relevance}.
To improve the sustainability of smart ecosystems, proper evolution mechanisms are required to be put in place. While evolution has a substantial body of knowledge in model-driven software engineering~\cite{di_ruscio2011what,hebig2017approaches}, hybrid cyber-physical components of smart ecosystems, such as digital twins~\cite{kritzinger2018digital}, give rise to challenges traditional software engineering techniques fall short of addressing.

Digital twins are real-time, virtual representations of physical system components~\cite{kritzinger2018digital}.
They govern smart ecosystems and provide essential mechanisms and services to assess, simulate, and control the physical infrastructure of smart ecosystems for optimal behavior~\cite{michael2024explaining}. Thus, to ensure the technical sustainability of smart ecosystems, first, the technical sustainability of digital twins must be managed.
Changes in digital twins boil down to a heterogeneous set of components, including software, hardware, middleware, and IoT devices. The interdependency of concerns severely hinders the applicability of software engineering techniques and even challenges the very understanding of evolutionary needs.

To help software engineers apply their expertise in digital twin evolution scenarios, we provide a \textit{case-based demonstration of the 7R taxonomy} in this paper. The 7R taxonomy of digital twin evolution~\cite{david2023towards} defines seven elementary activities to support the technical sustainability of digital twins.

This paper is structured as follows.
In \secref{sec:case}, we elaborate on a case of an evolving smart ecosystem, driven by digital twin evolution.
In \secref{sec:action-points}, we recommend action points to apply the 7R taxonomy.
In \secref{sec:conclusion}, we draw the conclusions.
We provide background information about key concepts in sidebars.
\begin{figure*}[t]
\begin{sidebarbox}

\begin{subfigure}{0.48\textwidth}
\subsection*{The 7R taxonomy of digital twin evolution}
Taxonomies are a form of classification, aiming to systematically organize knowledge of a specific research field or problem. Classification of objects helps to understand the specific field and systematically treat a particular problem. The 7R taxonomy of digital twin evolution~\cite{david2023towards} identifies seven areas of action to react to the evolutionary needs of digital twins.
\end{subfigure}%
\hfill
\begin{subfigure}{0.48\textwidth}
\centering
\includegraphics[width=\textwidth]{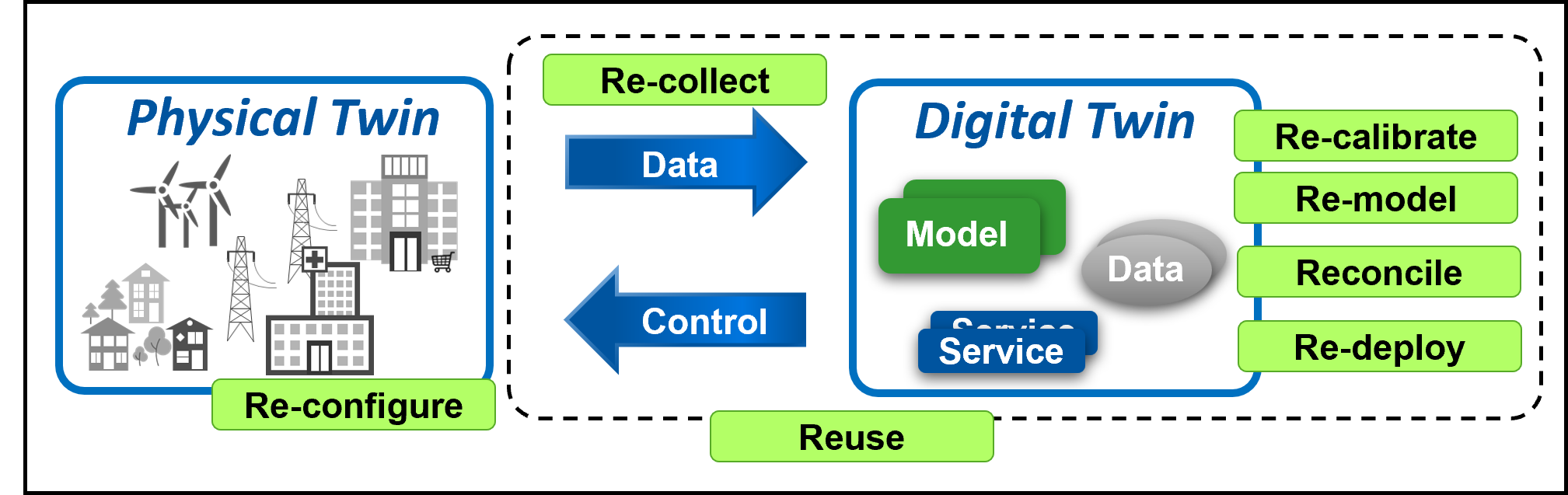}
\end{subfigure}

\phantom{}

\begin{subfigure}[t]{0.48\textwidth}
\textbf{Re-calibration} of a model parameter is required when the model is not a faithful representation of the physical twin anymore and simulations become incorrect, leading to imprecise assessment, analysis, and control of the physical twin.\\
\textbf{Re-modeling} the physical twin might be required in more elaborate cases, e.g., when the model does not reflect the real phenomenon properly. Specific software engineering tasks, such as re-architecting re-packaging a software component might be considered as refinements of this R-imperative.\\
\textbf{Reconciliation} of data, i.e., updating the data schema and migrating data might be needed when data discrepancies occur, and data might become inconsistent.
\end{subfigure}%
\hfill
\begin{subfigure}[t]{0.48\textwidth}
\textbf{Re-collecting} data is needed when events are missed due to transient errors. It might necessitate reconciliation, re-modeling, and re-calibration.\\
\textbf{Re-deploying} the evolved digital twin is needed after at least one of the previous steps has been taken.\\
\textbf{Re-configuration} of the physical twin is required after the digital twin has evolved. Re-configuration entails a wide range of potential actions, from changing the settings of a physical component to the installation of new ones.\\
\textbf{Reuse} of the large amounts of data, knowledge, and know-how that have been amassed during the operation of the digital twin is paramount in ensuring cost-efficient digital twin projects.
\addtocounter{figure}{-1}
\end{subfigure}
\end{sidebarbox}
\end{figure*}

\section{Case: Citizen Energy Community}\label{sec:case}

To illustrate the usage of the 7R taxonomy (see sidebar), we rely on a practical case of an evolving smart ecosystem, called the citizen energy community.

Energy communities enable collective, citizen-driven energy actions to support a clean energy transition~\cite{energy-communities}. In citizen energy communities (\figref{fig:usecase}), citizens and small commercial entities are equipped with \textit{energy generation} and \textit{storage capacity}, promoting them to first-class generators of energy. As opposed to traditional regulatory models, a citizen energy community gives rise to a \textit{smart ecosystem}, in which participation is voluntary and egalitarian; and cyber-physical components compose the infrastructure.

A digital twin is developed to govern the smart ecosystem~\cite{gramelsberger2023enabling} from the very beginning.
The digital twin provides stakeholders with tools to monitor and optimize energy trading processes, simulate energy provision and usage scenarios, analyze what-if scenarios, and predict maintenance requirements.

Throughout the lifespan of the system, new features are developed, new components are added, and core elements---often as critical as a power plant---are retired. In the following, we discuss four evolutionary scenarios in an escalating order of impact. By discussing the scenarios through the 7R framework of digital twin evolution for technical sustainability, we demonstrate how to organize the chain of thought about digital twin evolution into a structured set of arguments to support engineering tasks.

\begin{figure*}[t]
    \centering
    \includegraphics[width=\textwidth,trim={0 8cm 0 0},clip]{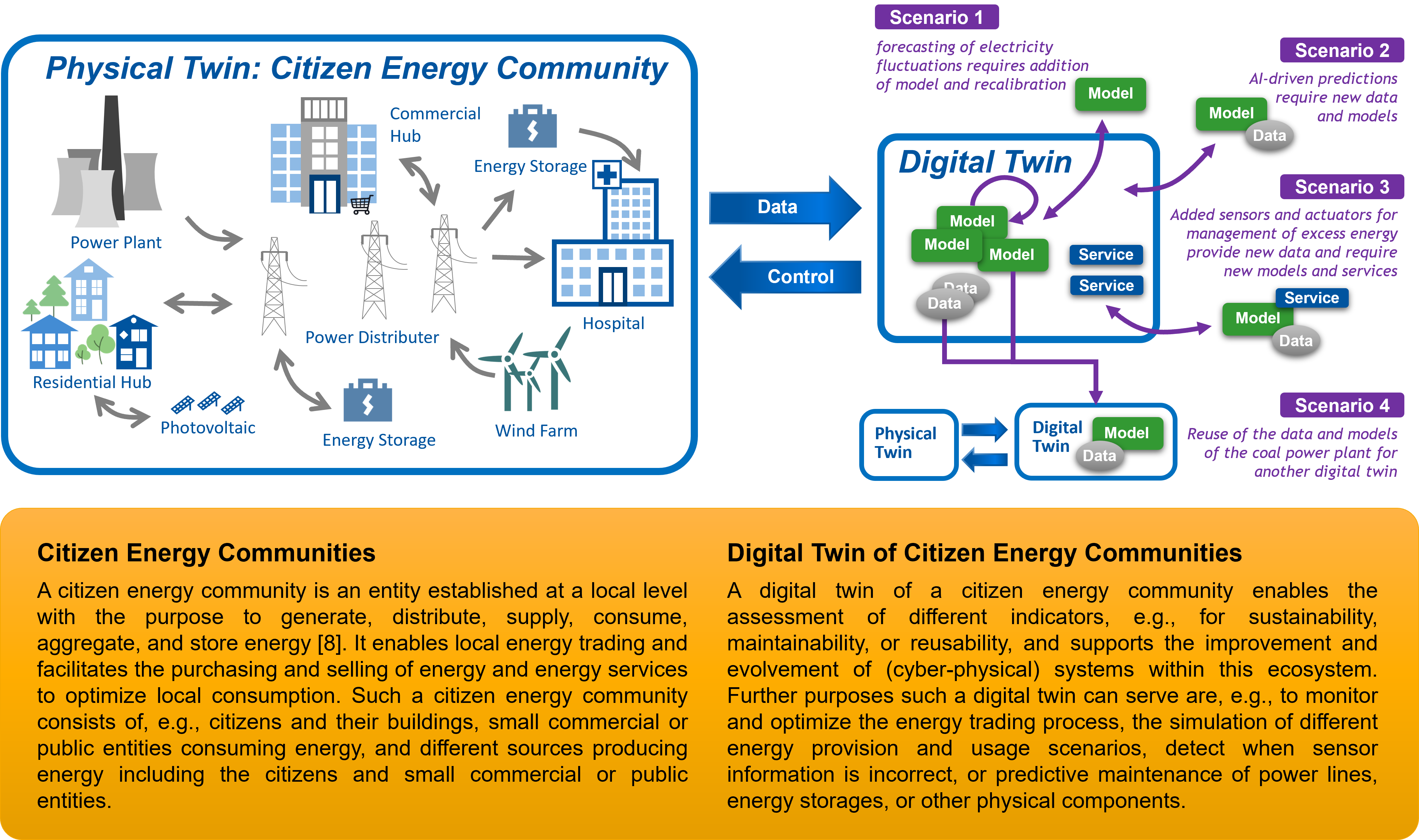}
    \vspace{-1.5em}
    \caption{Digital Twin of an Energy Citizen Community evolving over time}
    \label{fig:usecase}
\end{figure*}

\begin{figure*}[t]
\begin{sidebarbox}
\begin{subfigure}[t]{0.48\textwidth}
\subsection*{Citizen energy communities}
A citizen energy community~\cite{energy-communities} is a localized entity, established with the purpose of generating, distributing, supplying, and storing energy. It enables local energy trading and facilitates the purchasing and selling of energy and energy services to optimize local consumption~\cite{gramelsberger2023enabling}. Such a citizen energy community consists of citizens, their buildings, small commercial or public entities consuming energy, and different sources producing energy including the citizens and small commercial or public entities.
Energy communities are crucial in driving the clean energy transition. 
\end{subfigure}%
\hfill
\begin{subfigure}[t]{0.48\textwidth}
\subsection*{Digital twins of citizen energy communities}
A digital twin of a citizen energy community provides a faithful virtual replica of the overall socio-techno-economic system. By that, the digital twin enables the assessment of key indicators, e.g., of sustainability and overall system health, and supports the continuous improvement and evolution of the ecosystem. A digital twin also helps monitor and optimize energy trading processes~\cite{tsado2022digital}, simulate energy provision and usage scenarios, detect incorrect sensor information, and predict maintenance tasks of power lines, energy storages, or other physical components.
\end{subfigure}
\end{sidebarbox}
\addtocounter{figure}{-1}
\end{figure*}

\subsection{Scenario 1: From a monitoring digital twin to a predictive digital twin}\label{sec:case-scenario-1}

The local government decided to provide financial incentives to residents, who provide the excess energy of their photovoltaic systems within the citizen energy community network. In the new setting, client end-points do not only consume but also produce electricity. However, this setup necessitates accurate forecasting of electricity fluctuations, especially excess electricity to prevent damage, e.g., due to overheating components.

\subsubsection*{Re-model} Forecasting excess electricity requires a suitable model of the electrical grid. Engineering models that leverage laws of physics are a typical choice. Thus, the grid operator decides to improve the models of the digital twin and re-model the grid by adding models of thermodynamics and external factors, such as atmospheric pressure and relative humidity.

\subsubsection*{Re-calibrate} With the new models added, the digital twin needs to be re-calibrated. Without calibration, the models would not match the real system, resulting in inaccurate forecasts. Re-calibration is achieved by manual tuning based on high-quality operational data collected by the digital twin.

\vspace{-5pt}

\subsection{Scenario 2: AI-driven predictions}\label{sec:case-scenario-2}

After realizing the benefits of a predictive digital twin---e.g., improved resource efficiency and safety---the grid operator decides to further improve the predictive capabilities of the digital twin. One problem with the engineering model-based techniques in place is the computing power they require for detailed simulations. As an alternative, AI-based predictive methods are proposed and realized.

\subsubsection*{Re-collect} The development of the new AI model requires large volumes of data, including data that has not been considered before. Typically, data points that were excluded from the manually-built engineering models due to increased complexity are now becoming of particular interest, such as environmental data (e.g., cloud cover). Therefore, the data collection strategy needs to be revised, and the digital twin should start harvesting the required data points.

\subsubsection*{Reconcile} Collected data needs to pass through various data processing pipelines aiming to clean and consolidate data and eventually store it in a database. The data management infrastructure needs to be reconciled with the newly collected data. This includes technical aspects (e.g., updating data schemas and processing scripts); and in some cases, addressing the organizational or legal framework (e.g., when working with personal or sensitive data).

\subsubsection*{Re-model} After reconciliation, re-modeling is required to generate AI-based prediction models that are trained on data from the new data pipelines. The re-modeling, here, concerns the addition of new data quantities and qualities to establish an adequate model for predicting the behavior of the energy community using AI.

\subsubsection*{Re-calibrate} The evolution of the data and the model require a re-calibration of the model to adjust it to the evolved (i.e., extended) scope, end eventually, again faithfully reflect the physical twin.

\subsection{Scenario 3: Management of excess energy}\label{sec:case-scenario-3}

Too much energy can lead to voltage frequency disturbances in the system. As a result, transformers might trip off to protect themselves from being damaged. This can cause localized blackouts.
To further improve the safety of the grid and optimize its efficiency, the operator decides to equip the grid with the latest generation of safety components---sensors that detect potentially hazardous patterns, and actuators that can act upon hazardous situations. As usual, the digital twin operates these components.

\subsubsection*{Re-configure} First, the physical infrastructure of the grid needs to be re-configured. This re-configuration concerns putting new sensors and actuators in place. The new equipment enables the grid operator to localize causes for inefficient use of the grid and, consequently, to also actuate on identified grid components (e.g., temporal removal of consumers/producers from the grid, or establishment and enforcement of bandwidth limits).

\subsubsection*{Re-collect} As new sensors are in place that are producing data not considered before, the digital twin has to collect these new data points about hazardous situations such as voltage frequency disturbance or energy overload in specific areas of the grid. 

\subsubsection*{Re-model}  For the optimization of the smart grid efficiency, the operators decide to use the existing sensor and actuator components and integrate them to realize an agent who is in continuous interaction with the physical components by an actuation and sensing relationship. In this respect, a new model is created that supports a reinforcement learning approach~\cite{tomin2020development}.

\subsubsection*{Re-calibrate} The new model in support of reinforcement learning needs to be calibrated. This ensures that the model is a faithful representation of the grid. Calibration is achieved step-wise, by ingesting pieces of data as they arrive on the data stream.

\subsubsection*{Re-deploy} The data from the added sensors and actuators as well as the results of the developed reinforcement learning approach should be visualized to the users of the digital twin. This requires that the digital twin as a software system has to be re-deployed. 

\subsection{Scenario 4: Retiring the coal power plant}

Eventually, the distributed citizen energy community reaches the level of self-sustainability, efficiency, and safety, where the central coal power plant component is not needed anymore; and political trends drive the obsolescence of coal-fired power generation. As a consequence, the coal power plant is retired. The digital twin, however, is a source of important information thanks to the data collected throughout the lifespan of the coal power plant.
Additionally, legal constraints require the grid operator to keep this data for several years for documentation purposes.

\begin{table*}[t]
\centering
\small
\caption{Primary roles of Software Engineering in Digital Twin Evolution}
\label{tab:se-concerns}
\begin{tabular}{@{}lc>{\centering\arraybackslash}p{13cm}>{\centering\arraybackslash}p{1cm}@{}}
\toprule
\multicolumn{1}{c}{\multirow{2}{*}{\textbf{R-imperative}}} & & \multicolumn{2}{c}{\textbf{Involvement of software engineers}} \\
 & & \multicolumn{1}{c}{\textbf{Primary role}} & \multicolumn{1}{c}{\textbf{Extent}} \\ \cmidrule{1-1} \cmidrule{3-4}
Re-calibrate && Update models. In major cases: support model engineers and scientists. & \mediumSE \\
Re-model && Support model engineers and scientists, and refactor models for scalability. & \smallSE \\
Re-collect && Integration with sensor APIs and middleware (e.g., messaging). & \lotSE \\
Reconcile && Maintenance of data management pipelines, ETL processes, data schemas. & \lotSE \\
Re-deploy && Infrastructure-as-Code, DevOps, CI/CD. & \fullSE \\
Re-configure && Middleware development, embedded software development. & \mediumSE \\
Reuse && Software componentization for reuse. Transfer learning from AI components. & \smallSE\\

\bottomrule
\end{tabular}
\end{table*}

\subsubsection*{Reuse}
The grid operator is now able to reuse important design documents, design rationale (engineering decisions), experimental simulation traces, and operative information collected by the digital twin during the lifespan of the original power plant.
However, effective reuse might require further actions, e.g., re-calibrating models or re-collecting additional data.

Here, we maintain a focus on software aspects. In a system-wide focus, resource value retention options would become additionally important~\cite{david2024circular,bork2024role}, e.g., reusing particular components of a power plant, repairing or replacing parts in the smart grid, or re-purposing buildings leading to changed energy needs.
\section{Action points for application}\label{sec:action-points}

We aim to ease the application of the 7R taxonomy for digital twin evolution. Generally, applying the taxonomy requires answering two questions related to the affected R-imperatives on the one hand and the existing evolutionary processes on the other.

\subsection{Which of the R-imperatives does an evolutionary scenario touch upon?}

Answering this question helps in understanding the primary roles of software engineering in support of digital twin evolution, and the extent to which software engineering is involved in these phases. \tabref{tab:se-concerns} provides typical examples of such roles to every R-imperative.

\subsubsection*{Re-calibration} This imperative often does not require the involvement of model engineers and scientists; software engineers who are familiar with the model might take care of re-calibration in their own scope. Calibration and re-calibration of models is a moderately software-intensive R-imperative (\mediumSE[\inlineSize]).

\subsubsection*{Re-modeling} This imperative, on the other hand, is primarily the concern of model engineers and scientists. The role of software engineers is to take such models and refactor them for scalability. This is typical, e.g., with machine learning models, in which algorithms are fine-tuned by scientists, enabling software engineers to integrate the model into the software architecture. Re-modeling is one of the least software-intensive R-imperatives (\smallSE[\inlineSize]).

\subsubsection*{Re-collecting} Re-collecting data typically requires working with device APIs or interacting with a messaging middleware. It is a fairly software-intensive imperative (\lotSE[\inlineSize]) that touches upon distributed components and often runs into testing challenges.

\subsubsection*{Reconciliation} The software engineering effort focuses on maintaining data management pipelines as the underlying data collection infrastructure changes. This is a fairly critical and software-intensive imperative (\lotSE[\inlineSize]), as it touches upon data, a key value driver for companies~\cite{laney2017infonomics}.

\subsubsection*{Re-deployment} This imperative is typically the most software en-gineering--intensive one (\fullSE[\inlineSize]). As computing is typically located in the cloud nowadays, software engineers need to define the overall infrastructure-as-a-code~\cite{staron2023recent} for deployment, as well as enact the end-to-end DevOps or, in rare cases, CI/CD processes.

\subsubsection*{Re-configuration} Re-configuration of the physical infrastructure mostly requires interacting with middleware as physical components are mostly hidden behind messaging and procedural layers. Occasionally, developing and maintaining embedded software for physical devices might be required, which is typical in specialized cases, e.g., where custom measurement equipment is used. Still, this imperative is only moderately software-intensive (\mediumSE[\inlineSize]).

\subsubsection*{Reuse} This imperative can be supported by software engineering~\cite{MPRW22} by proper componentization of software, preparing it to be used in other digital twinning projects. AI-heavy companies might want to retain value from their previously trained AI components by transfer learning~\cite{farahani2020concise}. As reuse in digital twin settings is a more pressing challenge on the physical side of things, this R-imperative is one of the least software-intensive tasks (\smallSE[\inlineSize]).

\subsection{What are the processes in the organization?}

Answering this question helps organize the R-imperatives into a coherent flow. Taxonomies only define a classification of concepts and defer the operationalization to the specific context of the organization or company.
Thus, a process model or DevOps variant~\cite{david2023susdevops-techreport} is required to operationalize the taxonomy.
These operationalizations might differ in their extent, intent, and vendor dependence.

\subsubsection*{Extent: short versus long loops.}
In the demonstrative case, Scenario 1 is a relatively short loop. It requires implementing a new model and re-calibrating it. In contrast, Scenario 3 is a more elaborate one, touching upon all but one R-imperative. Clearly, the shorter the loop, the easier it is to oversee and manage. Evidence from the industry also shows that shorter loops, especially on the digital side of things (i.e., touching upon re-modeling, re-calibration, and re-deployment), are more frequently situated within the traditional realm of software engineering companies. Longer loops tend to extend into other domains and require more elaborate cooperation.

\subsubsection*{Intent: data-first versus model-first.}
In the demonstrative case, we show one particular sequence of R-imperatives for each scenario. In practice, R-imperatives can be chained in a different order and with more cycles to achieve the evolutionary goals of digital twins. Often, the preferred order of R-imperatives depends on company best practices and employed paradigms.

\begin{figure}[h]
    \centering
    \includegraphics[width=\linewidth]{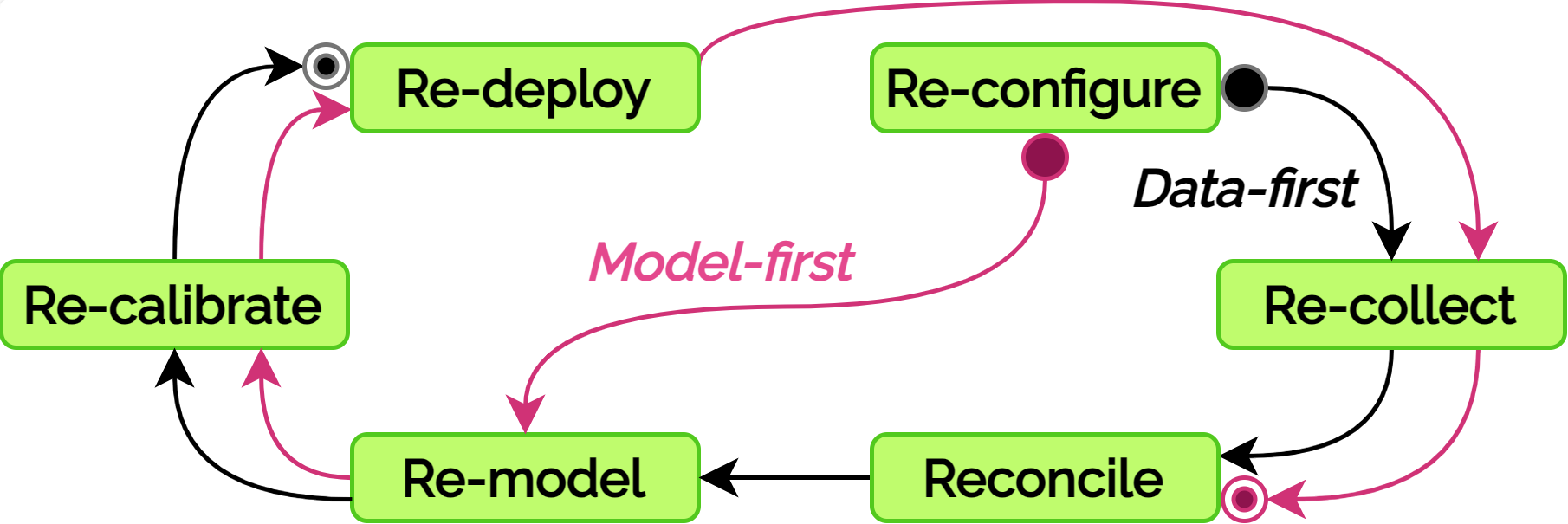}
    \caption{Operationalizations of the taxonomy in Scenario 3}
    \label{fig:operationalization}
\end{figure}

\figref{fig:operationalization} shows two typical operationalizations of Scenario 3. In a \textit{data-first} approach, the physical twin is re-configured, and subsequently, data collection and reconciliation start immediately to drive model creation in a deductive fashion. The discussion of Scenario 3 in the running example followed a data-first view. Alternatively, in a \textit{model-first} approach, the re-configuration of the physical twin is followed by re-modeling, re-calibration, and re-deployment of the digital twin. The benefit of this approach is that models can be used to re-generate data schemas and processing scripts, and thus, data collection can commence smoothly, almost without manual intervention. Software companies adopting model-driven practices~\cite{schmidt2006model} might venture into model-first evolutionary processes, but the data-first mindset is still prevalent in practice.

\subsubsection*{Vendor dependence.} Operating smart ecosystems is seldom a one-person show. Software companies work with various vendors. Increasingly more often, equipment vendors ship devices coupled with models pre-configured with reasonable defaults. In such cases, longer loops are to be expected, and re-modeling, re-calibration, and re-configuration tasks, in particular, need to be scheduled appropriately. In contrast, internal re-modeling and re-calibration speed up the process but pose challenges in technical aspects, such as maintenance, and non-functional aspects, such as certification.
\section{Conclusion}\label{sec:conclusion}

This paper provides a case-based introduction to the application of the 7R taxonomy of digital twin evolution. We focus on the role of software engineering in the key tasks outlined by the taxonomy (i.e., its R-imperatives).

Ultimately, the 7R taxonomy of digital twin evolution fosters better decisions in a convoluted problem space in which software engineers are key to success. There are many benefits software engineers can gain from using the taxonomy.

\subsubsection*{Comprehensive arguments} Scenarios can be explained in a structured way through the taxonomy, leading to more comprehensive arguments. The taxonomy defines the main concerns of digital twin evolution, and by explaining the scenarios through these concerns, the engineers can make sure they think of every important aspect.
    
\subsubsection*{Understanding the impact of change} The length of loops is indicative of the complexity of change and how demanding the evolution of the digital twin is. However, more comprehensive and structured impact analysis methods are required, given the critical nature of smart ecosystems. In the interim, we recommend relying on established cause-effect modeling techniques.
    
\subsubsection*{Mostly manual evolution} Mostly due to the physical components. Scenario 1 could have been addressed by putting the right change mechanisms in place. However, Scenario 2 showed that evolution can quickly escalate into the physical realm. Thus, better automation support for digital-physical co-evolution is required.

\bibliographystyle{ACM-Reference-Format}
\bibliography{bib/references}

\end{document}